# Transmission electron microscopy investigation of segregation and critical floating-layer content of indium for island formation in InGaAs

D. Litvinov and D. Gerthsen

*Laboratorium für Elektronenmikroskopie and Center for Functional Nanostructures (CFN), Universität Karlsruhe, D-76128 Karlsruhe, Germany*

A. Rosenauer and M. Schowalter

*Institut für Festkörperphysik, Universität Bremen, D-28359 Bremen, Germany*

T. Passow, P. Feinäugle, and M. Hetterich

*Institut für Angewandte Physik and CFN, Universität Karlsruhe, D-76128 Karlsruhe, Germany*

We have investigated InGaAs layers grown by molecular-beam epitaxy on GaAs(001) by transmission electron microscopy (TEM) and photoluminescence spectroscopy. InGaAs layers with In-concentrations of 16, 25 and 28 % and respective thicknesses of 20, 22 and 23 monolayers were deposited at 535 °C. The parameters were chosen to grow layers slightly above and below the transition between the two- and three-dimensional growth mode. In-concentration profiles were obtained from high-resolution TEM images by composition evaluation by lattice fringe analysis. The measured profiles can be well described applying the segregation model of Muraki et al. [Appl. Phys. Lett. **61** (1992) 557]. Calculated photoluminescence peak positions on the basis of the measured concentration profiles are in good agreement with the experimental ones. Evaluating experimental In-concentration profiles it is found that the transition from the two-dimensional to the three-dimensional growth mode occurs if the indium content in the In-floating layer exceeds $1.1\pm0.2$ monolayers. The measured exponential decrease of the In-concentration within the cap layer on top of the islands reveals that the In-floating layer is not consumed during island formation. The segregation efficiency above the islands is increased compared to the quantum wells which is explained tentatively by strain-dependent lattice-site selection of In. In addition, $In_{0.25}Ga_{0.75}As$ quantum wells were grown at different temperatures between 500 °C and 550 °C. The evaluation of concentration profiles shows that the segregation efficiency increases from $R=0.65$ to $R=0.83$.

# I. INTRODUCTION

InGaAs/GaAs-heterostructures have attained considerable interest due to numerous applications in (opto)electronic devices. Depending on the In-concentration, layer thickness and growth conditions, two-dimensional (2D) layers or three-dimensional (3D) islands are formed. Self-organized island formation occurs in the Stranski-Krastanov growth mode on an initial 2D wetting layer if the critical layer thickness is exceeded. The understanding of the factors which govern the 2D-3D transition has been the topic of numerous studies as reviewed e.g. by Shchukin et al. [1]. The critical thickness was associated recently with the amount of segregated indium on the growth surface. This motivates thorough studies of the In-segregation process in molecular-beam epitaxy (MBE) growth which is frequently applied to grow InGaAs/GaAs-heterostructures.

Several segregation models were proposed which allow the calculation of the In-concentration profile and the amount of segregated indium at the growth surface $x_s$. To describe segregation Moison et al. [2] suggested an exchange reaction of In and Ga between the surface and the underlying ("bulk") layer assuming thermodynamic equilibrium. Dehaese et al. [3] proposed a kinetic model involving a two-energy level system which leads to the same segregation effect as the Moison model for high growth temperatures above 500 °C, but describes additionally the kinetic limitation of segregation at low temperature (400 °C). Gerard et al. [4] showed that the validity of the Moison model is limited to In-concentrations below 11 %. Therefore, the model of Dehaese et al. should be also limited to In-concentrations below 11 % for high growth temperatures. Muraki et al. [5] suggested a phenomenological segregation model which implies that the In-concentration in the surface layer can exceed 1 monolayer (ML). The consequence of $x_s > 1$ ML must be that the indium in the surface layer is not fully incorporated in the crystal but is rather contained in a weakly bonded floating layer. Experimental evidence for such a layer was presented by Garcia et al. [6] and recently Martini et al. [7]. Experiments by Toyoshima et al. [8] showed that the 2D-3D growth-mode transition is correlated with the amount of In on the growth surface and occurs if $x_S$ exceeds 1.7 MLs. Based on a study of Walther et al. [9], Cullis et al. [10] developed a segregation-based model for the critical thickness of the 2D-3D transition. They suggested that the In-content in the floating layer governs the 2D-3D growth-mode transition independent of the nominal In-concentration of the bulk layer. This value was calculated to be 0.8 ML on the basis of the Dehaese model [3]. The critical In-content in the floating layer has to be clearly distinguished from the corresponding (critical) bulk layer thickness because the latter depends strongly on the In-concentration and the dependence of the In-segregation on the growth conditions.

Most experimental results regarding In-segregation were obtained by measuring $x_s$ in-situ in the MBE chamber by surface-sensitive techniques. However, there are still relatively few ex-situ studies which analyze the final In-concentration profiles, examples being transmission electron microscopy (TEM) studies the investigations of Walter et al. [8] and Rosenauer et al. [11,12], cross-section scanning tunneling microscopy studies [13-15] or secondary mass ion spectroscopy [5]. These ex-situ techniques are complementary to in-situ surface techniques which yield $x_s$ but not the final bulk concentration profiles.

In our TEM study we have measured quantitatively In-concentration profiles of a sample which was specifically grown to contain InGaAs layers slightly below and above the critical thickness for the 2D-3D transition. On the basis of measured In-concentration profiles, the segregation models suggested by Dehaese et al. [3] and Muraki et al. [5] were compared. This is particularly relevant with respect to the estimation of the critical In-content in the floating layer and the corresponding critical bulk layer thickness at which the 2D-3D growth-mode transition occurs. The measured In-concentration profiles are verified by the comparison of measured and calculated photoluminescence peak energies.

Additionally, we focus on In-segregation in and above capped islands and address the question of whether the In-floating layer is consumed by island formation or not. Since only information on the average In-concentration in electron-beam direction is available in a TEM experiment, we aimed at the formation of large islands in the growth experiment to limit the contribution of the embedding GaAs to the measured In-concentration profiles of islands. Therefore, a low In-concentration of 28 % was chosen to obtain a small lattice mismatch between islands and substrate which results in the formation of defect-free islands with a lateral size of approximately 40 nm. In addition, a series of samples was grown at different growth temperatures to derive the temperature dependence of the segregation efficiency which is the most relevant parameter for description of the segregation processes.

## II. EXPERIMENTAL PROCEDURES

The samples were grown by molecular-beam epitaxy on GaAs(001) substrates. After the growth of a GaAs buffer layer at 570 °C the substrate temperature was reduced to 535 °C. A 2x4 reconstruction is observed for the GaAs(001) surface during growth. For the first sample, in the following denoted as multilayer sample, three $In_xGa_{1-x}As$ layers were deposited which are separated by 28 nm thick GaAs layers. This structure was capped by 28 nm GaAs. The thicknesses and compositions of the $In_xGa_{1-x}As$ layers are 23, 20, and 22 monolayers and $x_0=0.28\pm0.03$ (layer 1), $x_0=0.16\pm0.02$ (layer 2), and $x_0=0.25\pm0.03$ (layer 3), respectively. The corresponding growth rates were 1.55 ML/sec, 1.33 ML/sec and 1.46 ML/sec. These values were determined *in-situ* by reflection high energy electron diffraction (RHEED) oscillations. Note that only during the deposition of the first InGaAs layer, a transition between 2D and 3D growth is observed by RHEED. The growth was interrupted between the three $In_xGa_{1-x}As$ layers to change the In cell temperature (i.e. the In-concentration). The As:Ga beam equivalent pressure ratio was 15:1 for the whole structure. $As_2$ source molecules were used. Five other samples were analyzed, each containing only one $In_xGa_{1-x}As$ quantum well with a nominal thickness of 22 ML and $x=0.25\pm0.03$, deposited at different growth temperatures between 500 °C and 550 °C.

The structural properties were studied by TEM of cross-section samples viewed along the [010]-zone axis, prepared by standard procedures [16]. Plan-view samples were prepared by chemical etching from the substrate side using a solution of NaOH (1 mol/l) and $H_2O_2$ (30%) with a proportion of 5:1 with the aim to prevent the formation of additional defects during preparation. For the TEM investigations a Philips CM 200 FEG/ST electron microscope with an electron energy of 200 keV was used. The microstructure of the plan-view and cross-section samples was analyzed by conventional TEM. The In-concentration in the InGaAs layers was obtained on an atomic scale by composition evaluation by lattice fringe analysis (CELFA) [17]. High-resolution TEM (HRTEM) lattice-fringe images were taken using [010] off-axis imaging conditions with a center of Laue circle (COLC) corresponding to (0,20,1.5) for imaging with the (002) reflection, whereas the COLC was (1.5,20,0) if the (200) reflection was used. Choosing a <100>-type zone-axis orientation is important because the amplitude of the chemically sensitive {200} reflections is strongly affected by the {111} reflections in a <110>-type zone axis due to dynamical electron diffraction and nonlinear image formation in TEM. For simplicity, the following description focuses on imaging with the (002) reflection. The chemically sensitive (002) reflection was centered on the optical axis. Only the (000) and (002) reflections were selected for the formation of lattice-fringe images. The local In-concentration was determined by measuring the amplitude of the (002) Fourier component of the image intensity. The local (002) amplitude is compared with (002) Fourier components calculated by the Bloch-wave method with structure factors which take also static atomic displacements into account [18]. Local thickness values of the TEM sample in regions with known composition, i.e. in the GaAs buffer layer adjacent to the islands, can be also obtained by the CELFA technique by using three-beam imaging conditions. More details of the evaluation procedure are outlined in Refs. [17,19].

Low temperature (5 K) photoluminescence spectra were acquired using an InGaAs detector and a spectrometer equipped with a 600 mm$^{-1}$ grating. The excitation was carried out by the 442 nm and 325 nm lines of a HeCd laser.

## III. EXPERIMENTAL RESULTS

Fig. 1 shows TEM images of the multilayer sample in a plan-view (Fig. 1a) and cross-section perspective (Figs. 1b,c). The inhomogeneous In-distribution in the islands, their sizes and 3D shape leads to the strain contrast in Fig. 1a where an arrangement of the islands mainly along the <100> directions can be seen. Figs. 1b,c are TEM dark-field images taken with the composition-sensitive (002) and (200) reflections. Figs. 1b,c reveal that the second and third InGaAs layers are 2D quantum wells (QWs) with a darker contrast compared to the embedding GaAs. The first InGaAs layer (bottom) with $x=0.28\pm0.03$ contains islands with sizes up to 40 nm and cores with a white contrast (Figs. 1b,c). The calculation of the image intensity of (002) dark-field images for InGaAs with the Bloch-wave method using structure factors computed within the density functional theory formalism [18] shows that minimum intensity in the InGaAs layer is observed if the In-concentration is approximately 17 %. The bright stripe in the center of the top InGaAs layer in Figs. 1b and the bright regions within the islands in the bottom layer indicate an In-concentration significantly larger than 17 % as expected for a nominal In-concentration $x=0.25\pm0.03$ and x=0.28±0.03.

Fig. 2 presents low-temperature (5 K) photoluminescence (PL) spectra of the multilayer sample, containing characteristic peaks for all the three layers which are labeled by the nominal In-concentration. Two sharp emission lines with a full width at half maximum of 8 and 7 meV occur at 1.315 meV and 1.385 eV. These lines can be attributed to the two InGaAs QWs. A broad emission band with a full width at half maximum of 77 meV is visible at approximately 1.18 eV which is in accordance with the observation of islands by TEM. A red shift of the PL peak positions with increasing nominal In-concentration is observed. In contrast to the narrow emission from the quantum wells, the width of the PL peak of the island layer is 10 times broader as a result of the distribution of island sizes and probably slightly different In-concentrations within the islands (see Fig. 1a).

For the composition determination in the multilayer sample, the CELFA technique is applied [17]. Fig. 3a shows a gray-scale coded map of the local In-concentration in the first InGaAs layer with islands (bottom) and the InGaAs quantum well (layer 2) with a nominal In-concentration of 16 %. Fig. 3a reveals that the In-concentration in the islands is strongly inhomogeneous with approximately 25 % at the bottom and 37 % close to the top. In Ref. [19] it was shown by simulated images that the CELFA evaluation of InGaAs QWs is rather insensitive to strain, lattice-plane bending, and inaccurately known values of specimen thickness and specimen orientation. However, in the present case we apply the CELFA technique to the investigation of islands, where the effect of strain and lattice-plane bending might be crucial. To investigate this influence, we compared CELFA evaluations of images formed with the (002) and (200) reflections. For the (002) lattice planes, which are parallel to the interface, the lattice distance within an island changes and the strain field induces lattice-plane bending. In contrast, the (200) lattice fringes (perpendicular to the interface) are only weakly affected by strain. As the islands are grown pseudomorphically, there is virtually no effect of strain on the (200) lattice planes in the center of a rotationally symmetric island. In the present case we found that concentration profiles obtained by CELFA with the (200) or (002) lattice planes give almost identical results. For the evaluation of segregation efficiencies, we prefer the (002) reflection, because the electron beam is (almost) parallel to the interface plane for a COLC of (0,20,1.5). Imaging with the (200) reflection induces a significant tilt of the electron beam with respect to the interface plane and broadening of the measured In-concentration profiles.

In-concentration profiles were obtained by averaging the measured In-concentration maps along the (horizontal) [100] direction. The error bars represent the standard deviation for the local In-concentrations along one monolayer which contains true concentration variations and variations due to noise on the image. Fig. 3b shows In-concentration profiles obtained for the layers 2 and 3 of the multilayer sample. The measured profiles were fitted with the segregation model of Muraki [5] using equation (1)

$$x(n) = \begin{cases} 0 & : \quad n < 1 \\ x_0(1-R^n) & : \quad 1 \leq n \leq N, \\ x_0(1-R^N)R^{n-N} & : \quad n > N \end{cases} \qquad (1)$$

where $n$ is the number of the ML in growth direction, $x_0$ is the nominal In-concentration, $R$ is the segregation efficiency and $N$ is the total amount of deposited In expressed in MLs of InGaAs. The parameters $x_0$, $N$ and $R$ were considered as fit parameters. As these parameters define different characteristics of the profile ($x_0$ and $R$ the slope of the increasing part and the maximum value, $R$ the shape of the decreasing part, and $N$ the position of the maximum), unique values of the fit parameters are obtained.

From the fitted profiles shown in Fig. 3b, In-segregation efficiencies of 0.80±0.01 for layer 2 and 0.79±0.01 for layer 3 are derived. Averaged values for $R$, which are obtained from different areas of the quantum well, yield $R$=0.81±0.02 for layer 2 and $R$=0.80±0.02 for layer 3. Note, that both layers were grown at the same temperature of 535 °C which allows the conclusion that $R$ is not affected by different values for $x_0$ (16 % and 25 % nominally) within the error limit. For the In-concentration $x_0$ we obtain (19±2) % for layer 2 and (25±2) % for layer 3.

Now we turn to the investigation of concentration profiles obtained for the islands (layer 1). Fig. 3c presents In-concentration profiles obtained from the center parts of two different islands. Due to the narrow region which was analyzed to derive the concentration profile error bars are not given because they would be unrealistically small. The profile labeled "island 1" corresponds to the left island in Fig. 3a, whereas an image of "island 2" is not presented here. The In-concentration profiles of the islands can be also well described by Eq.(1) which yields segregation efficiencies of $R$=0.84±0.01 for the island 1 and $R$=0.90±0.01 for the island 2. The island height is 32.7 and 30.6 ML, respectively. Evaluating several islands yields an average value $R$=0.86±0.04, which is slightly larger compared to the quantum wells (R ≈ 0.8).

## IV. DISCUSSION

**1. Description of In-segregation and evaluation of critical In-content for the 2D-3D growth mode transition**

The experimental results show (Fig. 3b) that the Muraki segregation model is well suited to describe the measured In-concentration profiles. We now compare the In-concentration profiles on the basis of the Muraki [5] and the Dehaese [3] models. The In subsurface/surface activation and segregation energies are taken from Ref. [3] to be 1.8 and 0.2 eV, together with the lattice vibration frequency of $10^{13}$ s$^{-1}$ (s. also [10]). Fig. 4a demonstrates that the Dehaese profiles deviate significantly from the Muraki profiles which model accurately the experimental profiles (Fig. 3b). This result is plausible, because the Dehaese model approaches the model of Moison et al. [2] at high growth temperatures above 500 °C, and Gerard et al. [4] showed that the Moison model is limited to In-concentrations below 11 %. For In-concentrations below 11 %, the segregation models of Muraki and Moison are in agreement [11]. Figure 4b depicts the amount of indium in the In-floating layer calculated according to the Muraki and Dehaese models for the layers 1 and 3. The amount of indium in the floating layer was given by Toyoshima et al. [8] for the Muraki model by Eq.(2):

$$FI(n) = x(n)\frac{R}{(1-R)} \qquad (2)$$

For layer 1, the nominal parameters were used. The maximum In-content in the floating layer calculated by the Muraki model is 0.99 ML for layer 3 and 1.11 ML for layer 1 whereas $x_s$ remains distinctly below 1 for the Dehaese model. As layer 1 is grown in the 3D mode in contrast to layer 3, we conclude that the critical amount of indium in the In-floating layer is 1.1±0.2. The error margin also takes the accuracy of the composition evaluation into account. This value is significantly smaller than the value 1.7 ML given by Toyoshima et al. in Ref. [7]. The discrepancy can be explained by two reasons. First, Toyoshima et al. used a modified version of the Muraki model (Eq. (3) in Ref. [8]), which yielded a slightly better fit of the experimental results, but at different segregation efficiencies. Using the same segregation efficiency, the modified Muraki model always gives larger In-contents in the In-floating layer. Second, the 2D-3D transition was obtained by RHEED, which could overestimate the critical thickness. Our value is somewhat higher than the value given by Cullis et al. [10]. This is the consequence of the application of the Dehaese model which tends to underestimate the In-content in the floating layer for high growth temperatures and In-concentrations above 11 % as also demonstrated in Fig.4b.

The segregation efficiency is the most relevant property for the quantitative description of the segregation process. It contains the effect of the growth conditions, in particular substrate temperature, V/III-flux ratio, surface reconstruction and the type of As-source molecule ($As_4$, $As_2$), which are known to influence In-segregation [20-22]. The temperature dependence of the In-segregation efficiency was investigated by analyzing a series InGaAs QWs with identical In-concentration (25 %), thickness (22 ML) and deposition conditions apart from the substrate temperature. The results for $R$ in the temperature range between 500 °C and 550 °C are shown in Fig. 5. The In-segregation efficiency increases significantly from 0.65 to 0.85, which demonstrates the strong dependence of R on the growth temperature in agreement with results of Kaspi et al. [23].

The In-concentration profiles for the QWs contained in the multilayer sample are verified by comparison of experimental and calculated PL peak energies. For that purpose we applied the envelope function formalism of Bastard [24] to the measured In-concentration profiles. The effective mass of electrons and holes are calculated in the vicinity of the Γ-point using a 3-band Kane model [25] and material parameters taken from Ref.26 are given in table 1. The Schrödinger equation was solved with a position-dependent potential deduced from the measured In-concentration profiles. The calculated PL at 5 K are 1.315 eV for layer 3 and 1.379 eV for layer 2. For the calculation of the concentration/potential profiles we used $R$=0.8, $N$=22 ML and $x_0$=0.25 for layer 3 and $R$=0.8, $N$=20 ML and $x_0$=0.19 for layer 2. Obviously, measured (1.315 eV) and calculated (1.315 eV) PL peak position are in good agreement for layer 3. For layer 2, the measured peak position was 1.385 eV deviates slightly from the calculated position of 1.379 eV. By variation of $x_0$ at constant values of $R$=0.8 and $N$=20 ML we obtain agreement for $x_0$=0.183 within the error limit of the value determined by TEM $x_0$= (0.19±0.02), which is also close to the nominal value of $x_0$=(0.16±0.02). In addition, nominal and measured thicknesses $N$ of layers 2 and 3 are in a good agreement.

## 2. In-segregation in islands

The evaluation of concentration profiles in the center of islands (layer 1, Fig. 3c) yields a segregation efficiency of $R$=0.86±0.04. This efficiency is slightly larger than the value obtained for the QWs grown at the same temperature. Additionally, values for $x_0$ of approximately 0.35 were determined, which is significantly larger than the nominal value of 0.28. These deviations can be explained by the following reasons.

One possible source of error is strain and lattice-plane bending. However, this effect can be ruled out, because concentration profiles observed in the center of the islands were similar for imaging with the (002) and (200) reflections.

Another point to consider is the fact that the islands are embedded in a GaAs matrix. In the TEM image we observe a projection of the island onto the viewing plane, which may contain contributions of an island and the surrounding cap layer. Fig. 6 shows a schematic illustration of plan-view and cross-section TEM samples with an island in the center. Typical lateral island sizes of approximately 40 nm are extracted from Fig. 1b,c. If the island is situated symmetrically within the TEM lamella with a thickness ≤ 15 nm, we observe an influence of the embedding matrix only in a small region close to the top of the island. This could explain the small dips in the In-concentration profiles and the deviations from the calculated segregation profiles in the upper part of the islands. For thicker TEM samples and asymmetrical island location in the sample, a reduced In-concentration will be measured.

In-concentrations above the nominal In-concentration within islands were also observed in Ref. [12] by TEM combined with strain state analysis. This effect was explained by strain-induced migration of In-atoms. Due to the elastic relaxation of an uncapped island during growth, In-atoms migrating along the growth surface prefer chemical bonding on top of the island due to the reduced strain energy compared to a site on top of the wetting layer. Careful measurement of In-composition distributions within uncapped islands with a nominal In-concentration of approximately 25 % grown at 540 °C were carried out by Walther et al. [9] using energy-selected imaging. These authors observed that the In-concentration of an island increases in growth direction and reaches approximately 60 % close to the top of the island.

Due to the lateral migration and accumulation of In-atoms on the growth surface above the islands, it appears questionable whether the In-concentration profile within an island can be described by the Muraki model for segregation, and there is no clear physical interpretation for the values of $x_0$ and $R$ obtained from the inner part of the islands as shown in Fig. 3c. On the other hand, the exponential decrease of the In-concentration observed within the cap layer clearly indicates the existence of an In-floating layer on top of the islands after the 2D-3D transition. The increased segregation efficiency above the islands reflects the affinity of the In-atoms to bond to lattice sites that minimize the strain energy due to the larger in-plane lattice parameter compared to the wetting layer between the islands. Therefore, regions of the growth surface on top of the islands are preferred sites for In-atoms, which effectively leads to the observation of an increased "segregation efficiency" compared to quantum wells. Our observations thus clearly demonstrate that the In-floating layer is not consumed by island formation.

## IV. SUMMARY

In this work we have measured In-concentration profiles of MBE-grown InGaAs quantum wells and islands capped with GaAs. The bulk In-concentration profiles are obtained by the evaluation of high-resolution TEM lattice-fringe images with the CELFA technique [17]. The experimental In-concentration profiles can be well described by the phenomenological segregation model of Muraki et al. [5], whereas significant discrepancies are encountered for the model suggested by Dehaese et al. [3] for our growth conditions. We obtain a segregation efficiency $R$=0.80±0.02 at a growth temperature of 535 °C for 2D layers with nominal In-concentrations of 0.16 % and 0.25 %. The analysis of the In-distribution in quantum wells with an In-concentration of 25 % shows that the segregation efficiency increases with growth temperature from 0.65 at 500 °C to 0.83 at 550 °C.

To determine the critical In-content in the In-floating layer for the 2D-3D growth mode transition, a sample was analyzed which contains InGaAs quantum wells and an island layer. On the basis of the measured segregation efficiency and the application of the Muraki model, which was confirmed explicitly by the experimental In-concentration profiles, we deduce a critical In-content of 1.1±0.2 ML.

In-concentration profiles in the center of capped islands reveal an exponential decrease of the In-concentration within the cap layer. Applying the segregation model of Muraki we find a segregation efficiency of 0.86±0.04 for regions on top of the islands. This result clearly shows that the In-floating layer on top of the islands is not consumed by island formation. The slightly larger segregation efficiency compared to the 2D layers is explained tentatively by the affinity of In-atoms to occupy lattice sites at the growth surface with minimum strain energy on top of the islands.

**ACKNOWLEDGEMENT**

This work has been performed within the project A.2 of the Deutsche Forschungsgemeinschaft (DFG) Research Center for Functional Nanostructures (CFN) at the University of Karlsruhe (Germany). It has been further supported by a grant from the Ministry of Science, Research and the Arts of Baden-Württemberg (Az: 7713.14-300).

# FIGURES

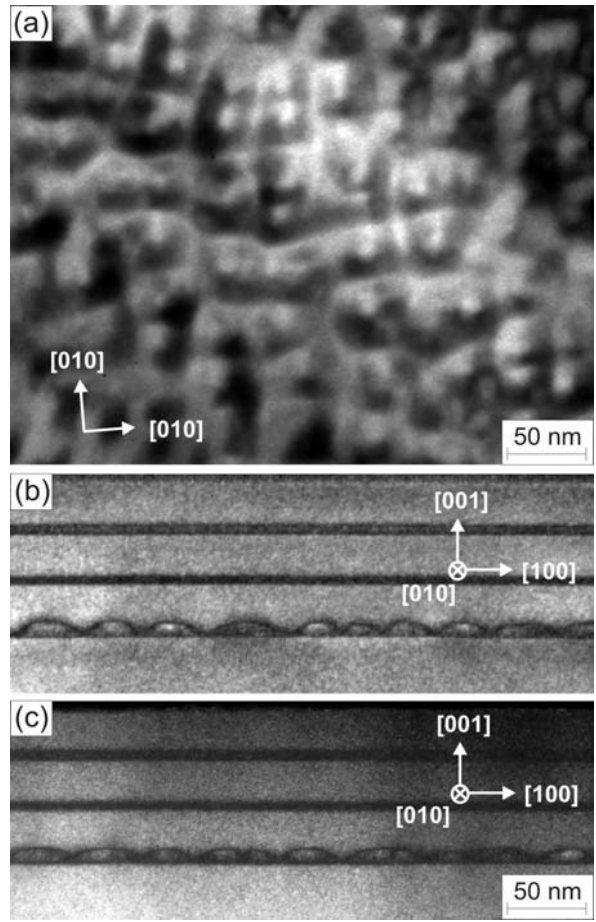

**Figure 1:** (a) Bright-field TEM plan-view image of the multilayer sample, (b,c) dark-field TEM images of a [010] cross-section of the multilayer sample with (b) **g** = (002) and (c) **g** = (200).

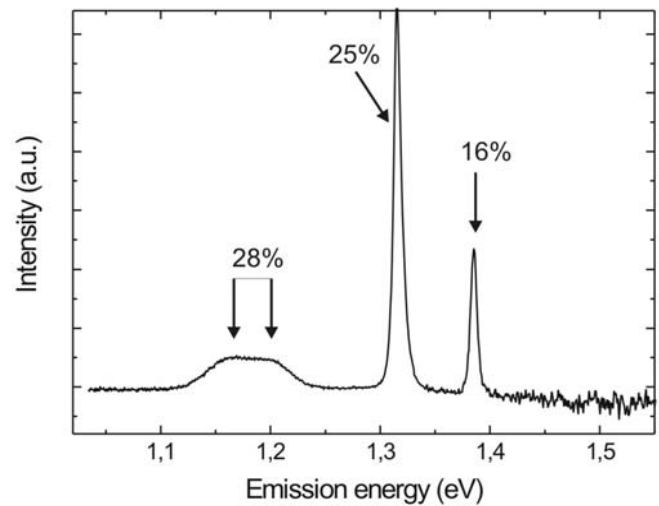

**Figure 2:** PL spectra of the multilayer sample recorded at 5 K. The three emission lines are labeled with the nominal In-concentration of the respective InGaAs layers.

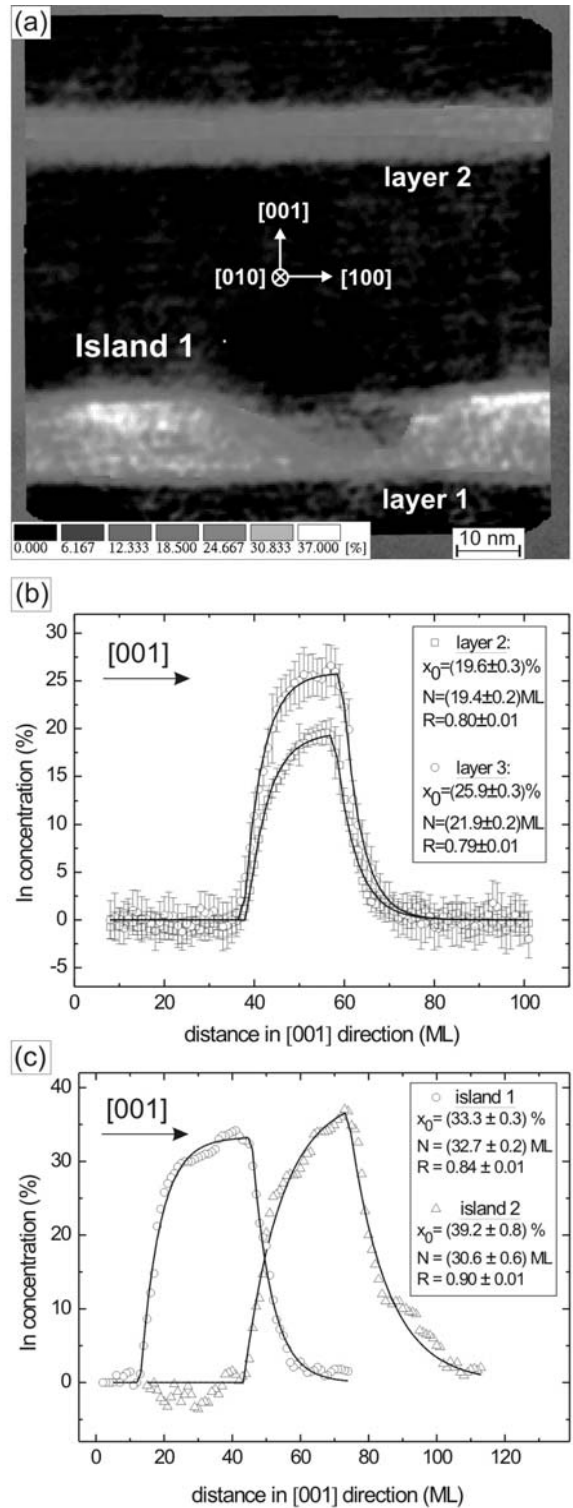

**Figure 3:** (a) Gray-scale coded map of the In-concentration in the first (bottom) and second (top) layers in the multilayer sample, obtained from a cross-section HRTEM lattice-fringe image with CELFA. (b,c) In-concentration profiles averaged along the [100] direction as a function of the distance in growth direction in units of ML (b) for layer 2 (squares) and layer 3 (circles) and (c) for the island 1 (circles). The triangles correspond to an island from another image not shown here. The solid curves are fit curves computed according to the segregation model of Muraki et al. [6]. The error bars give the standard deviation obtained by averaging along the respective lattice plane.

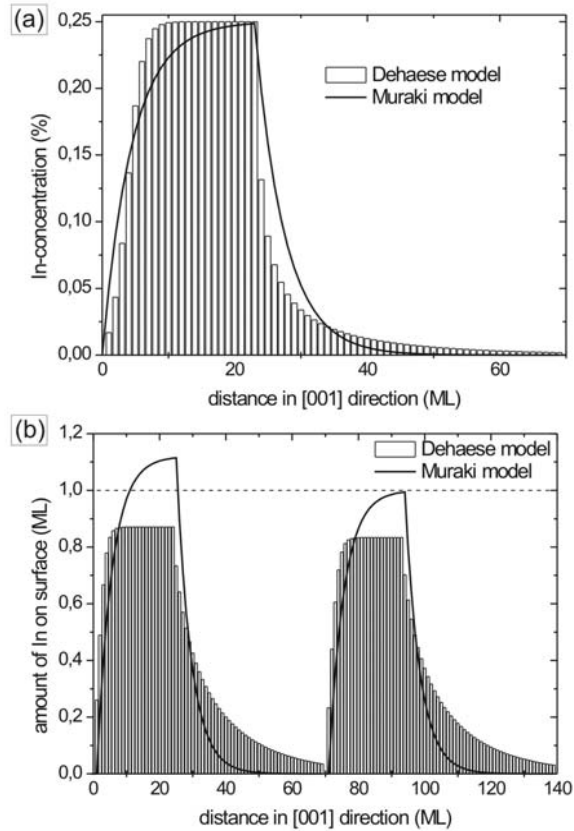

**Figure 4:** (a) Concentration profiles computed for layer 3 according to the segregation models of Muraki et al. [6] (solid line) and Dehaese et al. [12] (bars). For the Muraki model, we employed the parameters $R=0.8$, $N=22$ ML and $x_0=0.25$. The parameters for the Dehaese model were taken from Ref. [12]. (b) Amount of In in the In-floating layer during growth plotted vs. the distance in growth direction. For the calculation with the Muraki model, we used Eq. (2). The curves at the left-hand side were computed according to the nominal parameters of layer 1, whereas the curves at the right-hand side correspond to layer 3. As the 2D-3D transition is clearly exceeded in layer 1, we estimate that the critical amount of In lies close to 1 which is marked by dashed line.

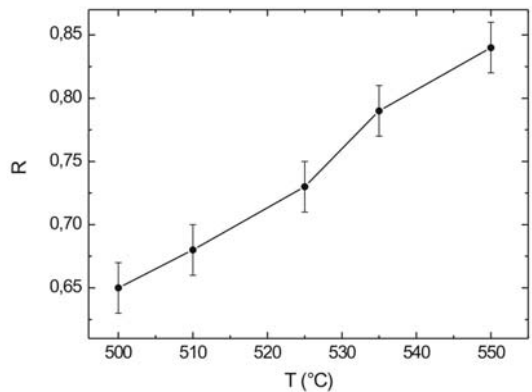

**Figure 5:** Segregation efficiency $R$ as function of the growth temperature $T$ for InGaAs QW layers with nominally $x_0=0.25$ and 22 ML thickness.

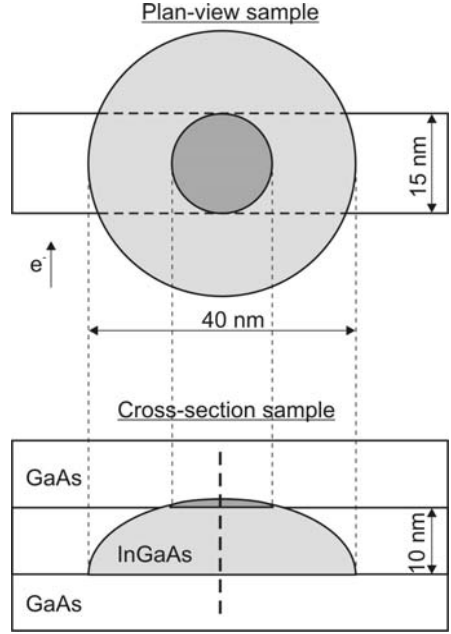

**Figure 6:** Schematic illustration of plan-view and cross-section samples with the island in the middle (dark region) for the discussion of the measured In-concentration profile for the island 1 in Fig. 3c.

**TABLE**

| Parameter | InAs | GaAs | Bowing parameter |
|---|---|---|---|
| Lattice parameter | 0.60583 nm | 0.56525 nm | |
| Energy gap | 0.4105 eV | 1.5192 eV | 0.475 eV |
| Varshni parameter $\alpha$ | 0.276 meV/K | 0.5405 meV/K | - |
| Varshni parameter $\beta$ | 93 K | 204 K | - |
| Effective mass $m_{eff}$ | 0.023 $m_0$ | 0.066 $m_0$ | - |
| Luttinger parameter $\gamma_1$ | 20.168 | 7.715 | - |
| Luttinger parameter $\gamma_2$ | 8.435 | 2.393 | - |
| Valence band offset | 0.62 eV | 0.62 eV | - |
| Deformation potential of conduction band | -4.91 eV | -7.7 eV | - |
| Deformation potential of valence band | -1.00 eV | -1.16 eV | - |
| Shear deformation potential | -1.80 eV | -1.7 eV | - |
| Elastic constant c11 | 833.0 GPa | 1188.0 GPa | - |
| Elastic constant c12 | 452.6 GPa | 538.0 GPa | - |

**Table 1:** Parameters used for the computation of the energy at maximum photoluminescence intensity. The parameters were taken from Ref.26. For the ternary material the parameter of the binary materials were linearly interpolated. A slight bowing was assumed for the energy gap.